\documentclass[preprint,showpacs,preprintnumbers,nobibnotes]{revtex4}
\usepackage{amsmath}
\usepackage{amssymb}
\usepackage{graphicx}
\usepackage{dcolumn}
\usepackage{bm}

\input{tcilatex}
\begin{document}

\title{\textbf{Phase diagrams and critical behavior of the quantum spin-}$%
1/2 $\textbf{\ XXZ model on diamond-type hierarchical lattices}}
\author{Xiu-Xing Zhang$^{1}$}
\author{Xiang-Mu Kong$^{1,2,}$}
\altaffiliation{Corresponding author. E-mail address: kongxm@mail.qfnu.edu.cn}
\author{Zhong-Yang Gao$^{1}$}
\author{Xiao-Song Chen$^{2}$}
\affiliation{$^{1}$Department of Physics, Qufu Normal University, Qufu, 273165, China\\
$^{2}$Institute of Theoretical Physics, Chinese Academy of Sciences, Beijing
100080, China}
\date{ \today }

\begin{abstract}
\ \ In this paper, the phase diagrams and the critical behavior of the spin-$%
1/2$ anisotropic XXZ ferromagnetic model (the anisotropic parameter $\Delta
\in \left( -\infty ,1\right] $) on two kinds of diamond-type hierarchical
(DH) lattices with fractal dimensions $d_{f}=2.58$ and $3$, respectively,
are studied via the real-space renormalization group method. It is found
that in the isotropic Heisenberg limit $\left( \Delta =0\right) $, there
exist finite temperature phase transitions for the two kinds of DH lattices
above. The systems are also investigated in the range of $-\infty <\Delta <0$
and it is found that they exhibit XY-like fixed points. Meanwhile, the
critical exponents of the above two systems are also calculated. The results
show that for the lattice with $d_{f}=2.58$, the value of the Ising critical
exponent $\nu _{\text{I}}$ is the same as that of classical Ising model and
the isotropic Heisenberg critical exponent $\nu _{\text{H}}$ is a finite
value, and for the lattice with $d_{f}=3$, the values of $\nu _{\text{I}}$
and $\nu _{\text{H}}$ agree well with those obtained on the simple cubic
lattice. We also discuss the quantum fluctuation at all temperatures and
find the fluctuation of XY-like model is stronger than the anistropic
Heisenberg model at the low-temperature region. By analyzing the
fluctuation, we conclude that there will be remarkable effect of neglecting
terms on the final results of the XY-like model. However, we can obtain
approximate result at bigger temperatures and give qualitatively correct
picture of the phase diagram.
\end{abstract}

\maketitle

\section{Introduction}

The spin-$1/2$ anisotropic Heisenberg model, or the XXZ model, is one of the
most important quantum spin models, which can be used to explain the quantum
essence of many magnetic materials, such as La$_{2}$CuO$_{4}$ and thin films
of $^{3}$He\cite{paper1}. Recently, for this model, many works have been
done on translational symmetric lattices\cite%
{paper2,paper3,paper4,paper5,squre,paper6,paper7}. For example, F C Alcaraz
and A L Malvezzi studied the XXZ chain in the presence of an external field
and obtained the phase diagram of this system accurately\cite{paper3}.
Continentino and Sousa, respectively, have investigated the critical
properties of the two-dimensional (2D) anisotropic Heisenberg model\cite%
{paper4,paper5,squre}, and found that, in the isotropic Heisenberg limit,
this system does not exhibit finite temperature phase transition (i.e., the
critical temperature $T_{C}=0$), which is in accordance with the theorem of
Mermin and Wagner\cite{paper8}. However, as to this model on the simple
cubic lattice $T_{C}$ no longer equals zero in the isotropic Heisenberg
limit, which means the existence of finite temperature phase transition\cite%
{paper6,paper7,paper9,paper10,paper11}.

On the other hand, there has been some interest in the critical phenomena of
XXZ model on fractal lattices, and particularly on the diamond-type
hierarchical (DH) lattices. Because of the special geometrical and
topological property, DH lattices are good candidates to investigate the
spin systems in non-integer dimensions, and since they have a much lower
symmetry than other fractals, so they may provide insights into other
low-symmetry problems such as random magnets, surfaces, and the like\cite%
{kaufman1,paper18}. In these aspects, some theoretical works have been done.
In 1983, using the renormalization group (RG) method, Caride \textit{et al}
have investigated the critical behavior of the Heisenberg model on the
Wheatstone-bridge-basis hierarchical lattice. Their results show that in the
isotropic Heisenberg limit $T_{C}$ approaches zero as a continuous function
of the anisotropic parameter\cite{paper12}. Latterly, the phase diagram of
this model was obtained by Souza on another kind of DH lattice with fractal
dimension $d_{f}=2$\cite{paper13}, and the results agree well with those on
the square lattice. Recently, using a real-space RG method, the anisotropic
Heisenberg spin-glass model on a three-dimensional DH lattice has been
studied\cite{Sousa}.

In the study of the anisotropic Heisenberg model, many effective approximate
methods have been applied. Such as the mean-field approximation\cite{paper9}%
, series expansion\cite{paper11}, RG method \cite{paper2,paper4,paper6} and
Monte Carlo simulation\cite{MonteCarlo,paper141}, etc. Among these methods,
the RG theory has been proved to be very powerful and it has been widely
used to investigate the critical behavior of different spin systems\cite%
{paper2,paper6,paper14,paper15}.

In this paper, using the RG method, we investigate the quantum spin-$1/2$
XXZ ferromagnetic model on two kinds of DH lattices with fractal dimensions $%
d_{f}=2.58$ and $d_{f}=3$, respectively. Our results show that the systems
exhibit finite temperature phase transitions in the isotropic Heisenberg
limit $(\Delta =0)$. Besides, we also investigate the above systems in the
range of $-\infty <\Delta <0$ and find that they exhibit XY-like fixed
points.

The outline of the remainder of this paper is as follows. In the next
Section, the model and the calculation method are presented; Sec. III gives
the results; Sec. IV is a discussion about some interesting quantum effects
and Sec. V gives a brief conclusion. Some of the more tedious of the
formulations in Sec. II are illustrated in the Appendix.

\section{Model and calculation method}

The effective Hamiltonian of the XXZ model can be written as\ \ \ \ \ \ 
\begin{equation}
H=K\sum_{\left\langle i,j\right\rangle }\left[ \left( 1-\Delta \right)
\left( \sigma _{i}^{x}\sigma _{j}^{x}+\sigma _{i}^{y}\sigma _{j}^{y}\right)
+\sigma _{i}^{z}\sigma _{j}^{z}\right] ,  \label{eq1}
\end{equation}%
where $K=J/k_{B}T$, in which $J$ is the exchange coupling parameter ( $J>0$
and $J<0$ correspond to the ferromagnetic and the antiferromagnetic model,
respectively), $k_{B}$ is the Boltzmann constant, and $T$ is the absolute
temperature. $\sigma _{i}^{\alpha }$ $\left( \alpha =x\text{, }y\text{, }%
z\right) $ are spin Pauli operators on the site $i$. The sum is over all the
nearest-neighbor spin pairs $\left\langle i,j\right\rangle $, and $\Delta
\in (-\infty $, $1]$ represents the anisotropic parameter. Note that the
Hamiltonian contains, as particular cases, the Ising model $($for $\Delta
=1) $, the isotropic Heisenberg model $($for $\Delta =0)$ and the XY model
(for $\Delta =-\infty $).

The DH lattices are all constructed by an iterative manner. Fig. 1$(a)$
illustrates the first three construction stages of DH lattice with fractal
dimension $d_{f}=2.58$ (lattice A, for simplicity). As can be seen, the
initiator is a two-point lattice joined by a single bond (construction stage 
$n=0$). Then the initiator is replaced with the generator ( the cluster of $%
n=1$ stage). Replacing every single bond on the generator itself, we get the
second stage of the lattice. If this procedure is repeated infinite times,
we then construct a DH lattice with self-similar structure\cite%
{paper16,paper162,paper17,paper18}. Using the same procedure as that of
lattice A, we can construct another kind of DH lattice (lattice B, for
simplicity) with fractal dimension $d_{f}=3$ (see Fig. 1$\left( b\right) $).

In this section, using the RG method proposed by Caride\cite{paper12,paper19}%
, we give the main calculation procedure of the XXZ ferromagnetic model on
lattice A. The case of lattice B can be solved in the same way.

In the next RG procedure, the Hamiltonians, except for the Ising limit $%
\left( \Delta =1\right) $ and the high temperature limit $\left( K=0\right) $%
, among the neighboring generators do not commute with each other, which
leads to the impossible of decoupling the genorator from the whole lattice.
In order to achieve this goal, the noncommutativity among the neighboring
generators were neglected. This method had been used in Refs.\cite%
{paper12,paper13,paper19,paper20} and the quantum effects and the
approximation of this method had been detailedly discussed in Ref.\cite%
{paper20}.

Based on the above approximation, we take out the generator from lattice A
to perform the RG transformation, which is shown in Fig. 2. As can be seen,
after summation of the internal spins $\sigma _{3}$, $\sigma _{4}$ and $%
\sigma _{5}$ (decimation), the generator (Fig. 2$(a)$) is transformed into a
new structure, which contains two spins (i.e., $\sigma _{1}$ and $\sigma
_{2} $) joined by a single bond (Fig. 2$(b)$). This procedure can be
described as%
\begin{equation}
\text{Tr}_{3,4,5}\exp \left( H_{13452}\right) =\exp \left( H_{12}^{\prime
}\right) ,  \label{eq2}
\end{equation}%
where $H_{13452}$ and $H_{12}^{\prime }$ are, respectively, the Hamiltonians
associated with the clusters $(a)$ and $\left( b\right) $ in Fig. 2, Tr$%
_{3,4,5}$ denotes the partial trace over states of the internal spins $%
\sigma _{3},\sigma _{4}$ and $\sigma _{5}$. According to Eq. (\ref{eq1}),
the Hamiltonians $H_{13452}$ and $H_{12}^{\prime }$ are%
\begin{eqnarray}
H_{13452} &=&K\left( 1-\bigtriangleup \right) [(\sigma _{1}^{x}\sigma
_{3}^{x}+\sigma _{1}^{y}\sigma _{3}^{y})+(\sigma _{1}^{x}\sigma
_{4}^{x}+\sigma _{1}^{y}\sigma _{4}^{y})+(\sigma _{1}^{x}\sigma
_{5}^{x}+\sigma _{1}^{y}\sigma _{5}^{y})  \notag \\
&&+(\sigma _{3}^{x}\sigma _{2}^{x}+\sigma _{3}^{y}\sigma _{2}^{y})+(\sigma
_{4}^{x}\sigma _{2}^{x}+\sigma _{4}^{y}\sigma _{2}^{y})+(\sigma
_{5}^{x}\sigma _{2}^{x}+\sigma _{5}^{y}\sigma _{2}^{y})]  \notag \\
&&+K\left( \sigma _{1}^{z}\sigma _{3}^{z}+\sigma _{1}^{z}\sigma
_{4}^{z}+\sigma _{1}^{z}\sigma _{5}^{z}+\sigma _{3}^{z}\sigma
_{2}^{z}+\sigma _{4}^{z}\sigma _{2}^{z}+\sigma _{5}^{z}\sigma _{2}^{z}\right)
\label{eq3}
\end{eqnarray}%
and%
\begin{equation}
H_{12}^{\prime }=K^{\prime }[(1-\bigtriangleup ^{\prime })(\sigma
_{1}^{x}\sigma _{2}^{x}+\sigma _{1}^{y}\sigma _{2}^{y})+\sigma
_{1}^{z}\sigma _{2}^{z}]+K_{0},  \label{eq4}
\end{equation}%
respectively, where $K_{0}$ is a constant included to make Eq. (\ref{eq2})
possible.

By calculating the partial trace in Eq. (\ref{eq2}), we can obtain the RG
recurrent relations between the new parameters $\left( K^{\prime }\text{, }%
\Delta ^{\prime }\right) $ and the original parameters $(K$, $\bigtriangleup
)$. Firstly, we expand $\exp \left( H_{12}^{\prime }\right) $ as%
\begin{equation}
\exp \left( H_{12}^{\prime }\right) =a^{\prime }+b_{12}^{\prime }(\sigma
_{1}^{x}\sigma _{2}^{x}+\sigma _{1}^{y}\sigma _{2}^{y})+c_{12}^{\prime
}\sigma _{1}^{z}\sigma _{2}^{z}.  \label{eq5}
\end{equation}%
Note that, in the expansion of $\exp \left( H_{12}^{\prime }\right) $, the
anticommutation among the spin Pauli operators are considered. Further, in
the basis which simultaneously diagonalize $\sigma _{1}^{z}$ and $\sigma
_{2}^{z}$, we express both sides of Eq. (\ref{eq5}) in the form of matrix.
The left-hand side of Eq. (\ref{eq5}) can be expressed as 
\begin{equation}
\exp \left( H_{12}^{\prime }\right) =\left( 
\begin{array}{cccc}
e^{\lambda _{1}^{\prime }} & 0 & 0 & 0 \\ 
0 & \frac{1}{2}(e^{\lambda _{2}^{\prime }}+e^{\lambda _{3}^{\prime }}) & 
\frac{1}{2}(e^{\lambda _{2}^{\prime }}-e^{\lambda _{3}^{\prime }}) & 0 \\ 
0 & \frac{1}{2}(e^{\lambda _{2}^{\prime }}-e^{\lambda _{3}^{\prime }}) & 
\frac{1}{2}(e^{\lambda _{2}^{\prime }}+e^{\lambda _{3}^{\prime }}) & 0 \\ 
0 & 0 & 0 & e^{\lambda _{4}^{\prime }}%
\end{array}%
\right) ,  \label{matrix6}
\end{equation}%
in which 
\begin{equation}
\lambda _{1}^{\prime }=\lambda _{4}^{\prime }=K^{\prime }+K_{0},  \label{eq7}
\end{equation}%
\begin{equation}
\lambda _{2}^{\prime }=-K^{\prime }+2W^{\prime }+K_{0}  \label{eq8}
\end{equation}%
and%
\begin{equation}
\lambda _{3}^{\prime }=-K^{\prime }-2W^{\prime }+K_{0}  \label{eq9}
\end{equation}%
are eigenvalues of $H_{12}^{\prime }$, where%
\begin{equation}
W^{\prime }=K^{\prime }\left( 1-\Delta ^{\prime }\right) .  \label{eq10}
\end{equation}%
Following the same steps as above, the right-hand side of Eq. (\ref{eq5})
can also be expressed in the form 
\begin{equation}
\left( 
\begin{array}{cccc}
a^{\prime }+c_{12}^{\prime } & 0 & 0 & 0 \\ 
0 & a^{\prime }-c_{12}^{\prime } & 2b_{12}^{\prime } & 0 \\ 
0 & 2b_{12}^{\prime } & a^{\prime }-c_{12}^{\prime } & 0 \\ 
0 & 0 & 0 & a^{\prime }+c_{12}^{\prime }%
\end{array}%
\right) .  \label{matrix11}
\end{equation}%
By combining Eqs. (\ref{matrix6})-(\ref{matrix11}), we get%
\begin{equation}
a^{\prime }+c_{12}^{\prime }=\text{exp}\left( K^{\prime }+K_{0}\right) ,
\label{eq12}
\end{equation}%
\begin{equation}
a^{\prime }-c_{12}^{\prime }=\frac{1}{2}\text{(exp}\left( K^{\prime
}-2K^{\prime }\Delta ^{\prime }+K_{0}\right) +\text{exp}\left( -3K^{\prime
}+2K^{\prime }\Delta ^{\prime }+K_{0}\right) \text{)}  \label{eq13}
\end{equation}%
and%
\begin{equation}
b_{12}^{\prime }=\frac{1}{4}\text{(exp}\left( K^{\prime }-2K^{\prime }\Delta
^{\prime }+K_{0}\right) -\text{exp}\left( -3K^{\prime }+2K^{\prime }\Delta
^{\prime }+K_{0}\right) \text{).}  \label{eq14}
\end{equation}%
Obviously, the coefficients $a^{\prime }$, $b_{12}^{\prime }$ and $%
c_{12}^{\prime }$ are all functions of $K^{\prime }$ and $\bigtriangleup
^{\prime }$.

So from Eqs. (\ref{eq12})-(\ref{eq14}), we obtain the relations between the
new parameters $\left( K^{\prime },\Delta ^{\prime }\right) $ and the
expansion coefficients $\left( a^{\prime },b_{12}^{\prime },c_{12}^{\prime
}\right) $ as follows 
\begin{equation}
\exp \left( 4K^{\prime }\right) =\frac{\left( a^{\prime }+c_{12}^{\prime
}\right) ^{2}}{\left( a^{\prime }-c_{12}^{\prime }\right)
^{2}-4b_{12}^{\prime 2}},  \label{eq15}
\end{equation}%
\begin{equation}
\exp \left( 4K^{\prime }\Delta ^{\prime }\right) =\frac{\left( a^{\prime
}+c_{12}^{\prime }\right) ^{2}}{\left( a^{\prime }-c_{12}^{\prime
}+2b_{12}^{\prime }\right) ^{2}}.  \label{eq16}
\end{equation}%
Analogously, $\exp \left( H_{13452}\right) $ can also be written as%
\begin{eqnarray}
&&\exp \left( H_{13452}\right) =a  \notag \\
&&+\sum_{\left\langle i,j\left( >i\right) \right\rangle }\left[ b_{ij}\left(
\sigma _{i}^{x}\sigma _{j}^{x}+\sigma _{i}^{y}\sigma _{j}^{y}\right)
+c_{ij}\sigma _{i}^{z}\sigma _{j}^{z}\right]  \notag \\
&&+\sum_{\left\langle i,j\left( >i\right) \right\rangle \neq \left\langle
k,l\left( >k\right) \right\rangle }\left[ d_{ij,kl}\left( \sigma
_{i}^{x}\sigma _{j}^{x}+\sigma _{i}^{y}\sigma _{j}^{y}\right) \sigma
_{k}^{z}\sigma _{l}^{z}+e_{ij,kl}\left( \sigma _{i}^{x}\sigma
_{j}^{x}+\sigma _{i}^{y}\sigma _{j}^{y}\right) \left( \sigma _{k}^{x}\sigma
_{l}^{x}+\sigma _{k}^{y}\sigma _{l}^{y}\right) \right]  \notag \\
&&+\sum_{\left\langle i,j\left( >i\right) \right\rangle \neq \left\langle
k,l\left( >k\right) \right\rangle }f_{ij,kl}\sigma _{i}^{z}\sigma
_{j}^{z}\sigma _{k}^{z}\sigma _{l}^{z},  \label{eq18}
\end{eqnarray}%
where $a$, $b_{ij}$, $c_{ij}$, $d_{ij,kl}$, $e_{ij,kl}$ and $f_{ij,kl}$ are
all functions of $K$ and $\bigtriangleup $, which can be determined by
diagonalizing both sides of Eq. (\ref{eq18}) in the same basis of $\sigma
_{1}^{z}$, $\sigma _{2}^{z}$, $\sigma _{3}^{z}$, $\sigma _{4}^{z}$ and $%
\sigma _{5}^{z}$. Because the matrices of the two sides are too lengthy, we
just give the diagonal matrix form of $H_{13452}$ in the Appendix.

From Eqs. (\ref{eq2}), (\ref{eq5}) and (\ref{eq18}), we can obtain the
relations among the expansion coefficients%
\begin{equation}
a^{\prime }=8a,  \label{eq19}
\end{equation}%
\begin{equation}
b_{12}^{\prime }=8b_{12}  \label{eq20}
\end{equation}%
and%
\begin{equation}
c_{12}^{\prime }=8c_{12}.  \label{eq21}
\end{equation}%
These expressions, together with Eqs. (\ref{eq15}) and (\ref{eq16}), we
finally get 
\begin{equation}
\exp \left( 4K^{\prime }\right) =\frac{\left( a+c_{12}\right) ^{2}}{\left(
a-c_{12}\right) ^{2}-4b_{12}^{2}},  \label{eq22}
\end{equation}%
\begin{equation}
\exp \left( 4K^{\prime }\Delta ^{\prime }\right) =\frac{\left(
a+c_{12}\right) ^{2}}{\left( a-c_{12}+2b_{12}\right) ^{2}}.  \label{eq23}
\end{equation}%
Since $a$, $b_{12}$ and $c_{12}$ are all functions of $K$ and $\Delta $, the
RG recurrent relations between the new parameters $(K^{\prime },\Delta
^{\prime })$ and the original parameters $(K,\Delta )$ are determined by
Eqs. (\ref{eq22}) and (\ref{eq23}). However, the analytical expression about
their right-hand sides are difficult to obtain. In order to obtain the phase
diagram and the fixed points, we will numerically solve the above two
equations and the results will be presented in the next section.

\section{Results}

By numerically iteration, the RG recurrent equations (\ref{eq22}) and (\ref%
{eq23}), the phase diagram and the fixed points for the system with fractal
dimension $d_{f}=2.58$ can be obtained, which are shown in Fig. 3. The curve 
$(a)$ in Fig. 3 gives the critical line of the system. As can be seen, in
the range of $\Delta \in \lbrack 0,1]$, the phase space is divided into the
paramagnetic phase (P) and the ferromagnetic phase (F) by the critical line.
In this range, we can also find that the system exhibits two unstable fixed
points, i.e., the Ising unstable fixed point (IP for short) $\left(
\bigtriangleup ,1/K\right) =$ $\left( 1,2.77\right) $ and the isotropic
Heisenberg unstable fixed point (HP) $\left( 0,1.55\right) $. In fact, the
value of the IP is in accordance with that of the classical model\cite%
{paper16,paper162}. At the IP, the correlation length critical exponent can
be calculated by 
\begin{equation}
\nu _{\text{I}}=\frac{\ln b}{\ln \lambda },  \label{eq24}
\end{equation}%
in which, $b=2$ is the scaling factor and $\lambda =\left( \partial
K^{\prime }/\partial K\right) _{\bigtriangleup =1,K=0.36}=1.86$, thus, $\nu
_{\text{I}}=1.12$. We note that this value is the same as that of the
classical Ising model\cite{paper16,paper162,paper17}. It is worth mentioning
that, at the HP, the system exhibits a finite temperature phase transition
with critical temperature $T_{C}=1/K_{C}=1.55$. However, this is disagree
with the result of the Wheatstone-bridge-basis hierarchical lattice\cite%
{paper12}, where only zero temperature phase transition exists, i.e., $%
T_{C}=0$ when $\Delta =0$ (curve $\left( b\right) $ in Fig. 3). In addition,
the isotropic Heisenberg critical exponent is obtained as $\nu _{\text{H}%
}=2.04$ which differs greatly from the results $(\nu _{\text{H}}=\infty )$
of the 2D regular lattices\cite{paper6,paper7} and other hierarchical
lattices with lower fractal dimensions $\left( d_{f}<2.58\right) $\cite%
{paper12,paper13}.

We also investigate the above system in the range of $-\infty <\Delta <0$.
Our results show that the system exhibits an unstable fixed point, which is
not given in Refs. \cite{squre,paper12} and we call it XY-like fixed point
(XYP). The critical line between the HP and the XYP divides the phase space
into paramagnetic phase (P) and ferromagnetic phase (F) as well.

Using the same calculation procedure as the lattice A, we can also
investigate the critical behavior of the XXZ model on the lattice B, i.e.,
the DH lattice with $d_{f}=3$. The phase diagram and the fixed points are
presented in Fig. 4. We can see that the phase space is divided into
paramagnetic phase (P) and ferromagnetic phase (F) by the critical line. In
the range of $\Delta \in \left[ 0,1\right] $, the system also exhibits two
unstable fixed points, i.e., the Ising fixed point (IP) and the isotropic
Heisenberg fixed point (HP). At the IP $\left( \bigtriangleup ,1/K\right) =$ 
$\left( 1,3.79\right) $ we obtain the critical temperature $T_{C}=3.97$ and
the critical exponent $\nu _{\text{I}}=0.95$. Compared with the results of
three dimentional Heisenberg model calculated by other RG methods\cite%
{paper20,paper21}, our results are more consistent with that from series
expansion ($T_{C}=4.54,\nu _{\text{I}}=0.63$)\cite{paper11}. At the HP $%
\left( \bigtriangleup ,1/K\right) =$ $\left( 0,2.36\right) $, the critical
temperature $T_{C}=2.36$ is very close to that of three-dimensional
Heisenberg system ($T_{C}=2.41$)\cite{paper10}.\textit{\ }In addition, the
critical exponent is $\nu _{\text{H}}=1.69$ which agrees well with that
obtained by Sousa \textit{et al }on the simple cubic lattice where $\nu _{%
\text{H}}=1.64$\cite{paper22}. In summary, we can see that the lattice B can
be regarded as an approximation for the simple cubic lattice. Besides, just
as the case of the lattice A, the system on lattice B also exhibits a
XY-like fixed point (XYP) $\left( \bigtriangleup ,1/K\right) =\left(
-1.99,8.44\right) $.

\section{Discussion}

In this section, we discuss the possible effects of quantum fluctuation. We
assume the generator of lattice A as a whole. In this case, there will be no
noncommutativity and the result should be exact in our calculation. With
this method, the exact calculation of the DH lattice will be noted $%
[K^{^{\prime }}(K,\Delta ),\Delta ^{^{\prime }}(K,\Delta )]$. On the other
hand, for the RG transformation in Fig. 2, we apply a modified
Migdal-Kadanoff method, which is an approximate one. In this method, the
original cell (Fig. 2$(a)$) can be considered as a combination of 3 arrays
in parallel, each of which is made up of two interactions in series. The
renormalized interaction $K$ and anisotropy $\Delta $ can be firstly
calculated for each combination in series and then combined in parallel. The
approximate calculation provides $[3K^{^{^{\prime \prime }}}(K,\Delta
),\Delta ^{^{\prime \prime }}(K,\Delta )]$. We use the convenient ratios
introduced in Ref. \cite{paper19}%
\begin{equation}
R^{K}=\frac{3K^{^{^{\prime \prime }}}(K,\Delta ;K,\Delta )}{K^{^{\prime
}}(K,\Delta )}  \label{eq25}
\end{equation}%
and%
\begin{equation}
R^{\Delta }=\frac{\Delta ^{^{\prime \prime }}(K,\Delta ;K,\Delta )}{\Delta
^{^{\prime }}(K,\Delta )}  \label{eq26}
\end{equation}%
The $T$-dependences of $R^{K}$ and $R^{\Delta }$ for typical values of $%
\Delta $ are indicated in Fig. 5. In the high temperature limit, both $R^{K}$
and $R^{\Delta }$ tend to unity for all values of $\Delta $ ( $0\leq \Delta
\leq 1$ and $\Delta <0$ ). In the range of low temperature, where the
quantum effect tends to drive the system to a disordered phase, both $R^{K}$
and $R^{\Delta }$ show oscillational behaviours. The fact is usually due to
quantum fluctuation which, at low temperatures ($T$) and smaller anistropy ($%
\Delta $) are important\cite{Sousa}. As we see in Fig. 5, the quantum
fluctuation of XY-like model ($\Delta <0$) is stronger than the anistropic
Heisenberg model ($0\leq \Delta \leq 1$) at the low-temperature region. So,
there will be considerable error when we calculate the XY-like model at
lower temperatures. However, in the range of $0\leq \Delta \leq 1$, it is a
very good approximation and when the anistropy $\Delta $ is negative, as we
can see in Fig. 5, it follows the same tendency as the positive one. So, we
can obtain approximate result at bigger temperatures and give qualitatively
correct picture of the phase diagram. When it comes to the lattice B, both $%
R^{K}$ and $R^{\Delta }$ have the same type of behaviour and we can obtain
the same conclusion.

\section{Conclusion}

In conclusion, using the RG method, we have studied the quantum spin-$1/2$
anisotropic XXZ model on two kinds of DH lattices with fractal dimensions $%
d_{f}=2.58$ and $3$, respectively. Phase diagrams, fixed points and critical
exponents of the systems are obtained. The results show that, in the Ising
limit $\left( \Delta =1\right) $, the values of the fixed points and the
critical exponents agree well with those in Refs. \cite%
{paper11,paper162,paper17}. In the isotropic Heisenberg limit $\left( \Delta
=0\right) $, there are finite temperature phase transitions on the above two
lattices. Furthermore, the systems exhibit XY-like fixed points in the range
of $-\infty <\Delta <0$. The quantum effects of this system show that, at
low temperatures, the XY-like model has stronger fluctuation than the
anistropic Heisenberg model. So, we conclude that there will be a
considerable error when we calculate the XY-like model at low temperature.
Besides, our results also indicate that the DH lattice with $d_{f}=3$ can be
regarded as an approximation for the simple cubic lattice. As a comment,
this method can be extended to investigate spin systems with $S>1/2$ on
other lattices and we are presently working along these lines.

\begin{acknowledgments}
This work is supported by the National Science Foundation for Post-doctoral
Scientists of China (2005037442) and the Science Foundation of Qufu Normal
University. We thank Prof. Shu Chen for useful discussions. One of the
authors (Zhang) would like to acknowledges many fruitful discussions with
Zhong-Qiang Liu, Guang-Hou Sun, Xian-Ming Li, Xun-Chang Yin, Xin Zhang and
Zhong-Yang Gao.
\end{acknowledgments}

{\Large Appendix}

In the basis of $\sigma _{1}^{z},\sigma _{2}^{z},\sigma _{3}^{z},\sigma
_{4}^{z}$ and $\sigma _{5}^{z}$, $H_{13452}$ can be expressed as a $32\times
32$ matrix in the form of

\begin{equation*}
H_{13452}=\left( 
\begin{array}{cccccc}
A & 0 & 0 & 0 & 0 & 0 \\ 
0 & B & 0 & 0 & 0 & 0 \\ 
0 & 0 & C & 0 & 0 & 0 \\ 
0 & 0 & 0 & C & 0 & 0 \\ 
0 & 0 & 0 & 0 & B & 0 \\ 
0 & 0 & 0 & 0 & 0 & A%
\end{array}%
\right) ,
\end{equation*}%
in which $A=6K,$

\begin{equation*}
B=\left( 
\begin{array}{ccccc}
0 & 0 & 2W & 2W & 2W \\ 
0 & 0 & 2W & 2W & 2W \\ 
2W & 2W & 2K & 0 & 0 \\ 
2W & 2W & 0 & 2K & 0 \\ 
2W & 2W & 0 & 0 & 2K%
\end{array}%
\right) ,
\end{equation*}%
\begin{equation*}
C=\left( 
\begin{array}{cccccccccc}
-6K & 2W & 2W & 2W & 2W & 2W & 2W & 0 & 0 & 0 \\ 
2W & 0 & 0 & 0 & 0 & 0 & 0 & 2W & 2W & 0 \\ 
2W & 0 & 0 & 0 & 0 & 0 & 0 & 2W & 0 & 2W \\ 
2W & 0 & 0 & 0 & 0 & 0 & 0 & 0 & 2W & 2W \\ 
2W & 0 & 0 & 0 & 0 & 0 & 0 & 2W & 2W & 0 \\ 
2W & 0 & 0 & 0 & 0 & 0 & 0 & 2W & 0 & 2W \\ 
2W & 0 & 0 & 0 & 0 & 0 & 0 & 0 & 2W & 2W \\ 
0 & 2W & 2W & 0 & 2W & 2W & 0 & -2K & 0 & 0 \\ 
0 & 2W & 0 & 2W & 2W & 0 & 2W & 0 & -2K & 0 \\ 
0 & 0 & 2W & 2W & 0 & 2W & 2W & 0 & 0 & -2K%
\end{array}%
\right) ,
\end{equation*}%
where $W=K\left( 1-\Delta \right) .$

\newpage

{\Large Figure Captions}

\bigskip

\bigskip

\bigskip

Fig. 1 \ The first two construction stages of the DH lattices. $(a)$ the DH
lattice with fractal dimension $d_{f}=2.58$, $(b)$ the DH lattice with $%
d_{f}=3$.

\bigskip

\bigskip

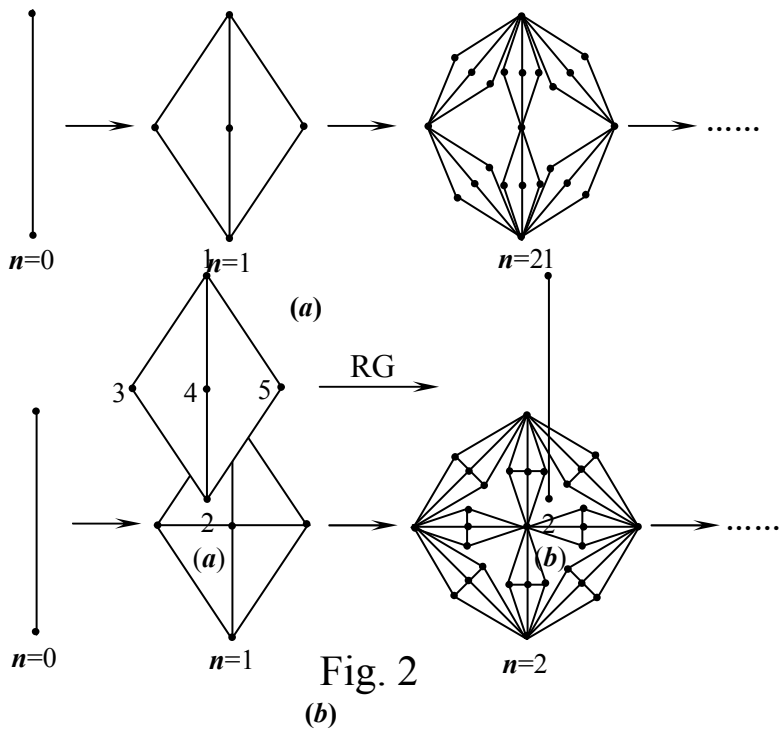
Fig. 2 \ The procedure of the RG transformation. After a step of the RG
transformation the generator $(a)$ is transformed into a bond $(b)$.

\bigskip

\bigskip

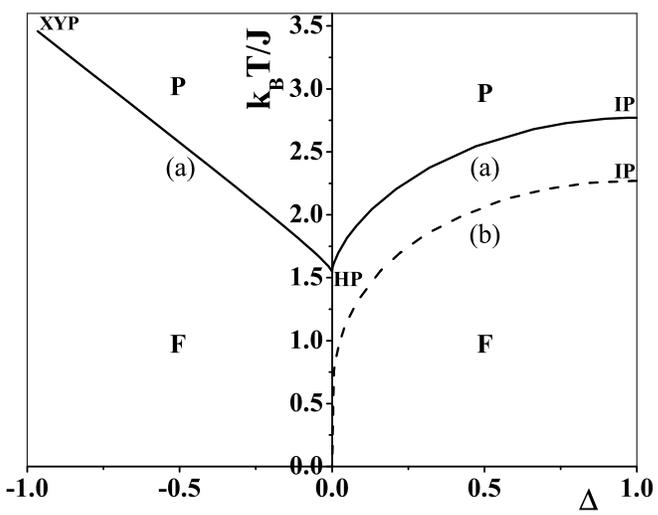
Fig. 3 \ Phase diagram of the DH lattice with $d_{f}=2.58$. The critical
line $\left( a\right) $ separates the phase space into paramagnetic phase
(P) and ferromagnetic phase (F). The IP, HP and XYP, respectively, denote
the Ising, isotropic Heisenberg and XY fixed points. $(b)$ is the critical
line of the Wheatstone-bridge-basis hierarchical lattice for comparison.

\bigskip

\bigskip

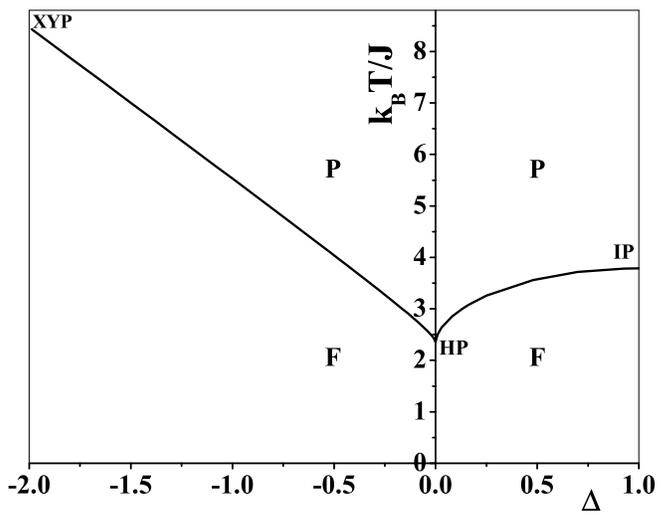
Fig. 4 \ Phase diagram of the DH lattice with $d_{f}=3$. The IP, HP and XYP,
respectively, denote the Ising, isotropic Heisenberg and XY fixed points. P
and F correspond to the paramagnetic phase and ferromagnetic phase,
respectively.

\bigskip

\bigskip

\bigskip

Fig. 5 \ Thermal dependence of the ratios $R^{K}$ and $R^{\Delta }$
respectively defined by Eqs. (24) and (25), for typical values of $\Delta $.

\bigskip

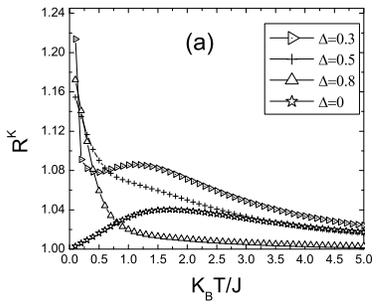
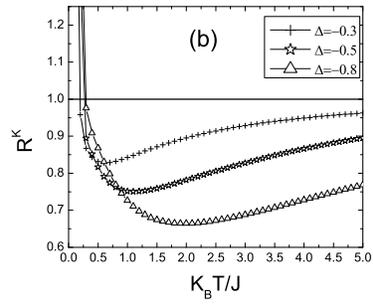
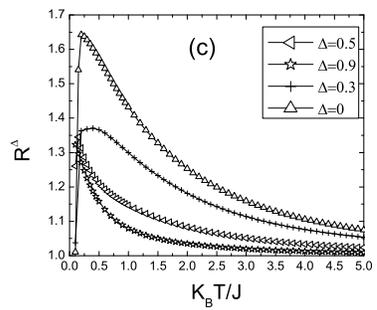
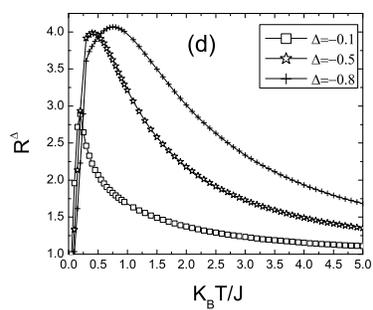
\end{document}